\pdfoutput=1
\documentclass[a4paper, 12pt]{article}

\usepackage{amsmath,amsthm}
\usepackage{parskip}
\usepackage{subfig}
\usepackage{authblk}
\usepackage[margin=1in]{geometry}
\usepackage{amssymb}
\usepackage{stmaryrd}
\usepackage{hyperref}
\usepackage{babel}
\usepackage{xcolor}
\usepackage{graphicx}
\usepackage{cancel}
\usepackage[normalem]{ulem}
\usepackage{babel}
\usepackage[numbers] {natbib}

\begin{document}

 \title{Explicit lump and line rogue wave solutions to a modified Hietarinta equation}
\author[1]{Solomon Manukure\footnote{solomon.manukure@famu.edu (Corresponding Author)}}
\author[2]{Morgan McAnally}
\author[3]{Yuan Zhou}
\author[4]{Demetrius Rowland}
\author[5]{Gina Pantano}
\affil[1]{\small Department of Mathematics, Florida A\&M University, Tallahassee, FL 32307, USA}
\affil[2,5]{\small Department of Mathematics, The University of Tampa, Tampa, FL 33606, USA}
\affil[3]{\small School of Business, Xianda College of Economics \& Humanities, Shanghai International Studies University, Shanghai, 200083, China}
\affil[4]{\small Department of Mathematics, The University of Texas at Austin, Austin, TX 78712, USA}
\date{}
\maketitle

 \begin{abstract}
{Lump solutions are spatially rationally localized solutions which usually arise as solutions to higher dimensional nonlinear partial differential equations often possessing Hirota bilinear forms. Under some parameter constraint, these solutions may lead to rogue wave solutions. In this article, we study lump and rogue wave solutions of a new nonlinear non-evolutionary equation in 2+1 dimensions with the aid of a computer algebra system. We present illustrative examples and analyze the dynamical behavior of the solutions using graphical representations.}\newline\\	
\textbf{Key words:} Lump solutions, Rogue waves, Hirota bilinear form, Hietarinta equation \\
\textbf{PACS codes:} 02.30.Ik, 04.20.Fy, 05.45.Yv \\
\textbf{MSC codes:} 35C08; 35Q51
  \end{abstract}
   
\maketitle
\section{Introduction}
Lump solutions, which are analytic and spatially localized rational solutions to higher dimensional nonlinear partial differential equations, have been an active research area in mathematical physics over the last few years. They were first found by Manakov and Zakharov \cite{manakov1977twodimensional} for the KP equation by taking long wave limits of $N$-solitons \cite{manakov1977twodimensional, satsuma1979two,liu2019lump,liu2020dynamics}, but have recently been found in several integrable and nonintegrable equations (see e.g., \cite{satsuma1979two,ma2015lump, manukure2018lump,ren2019characteristics,zhou2019lump,zhao2019rogue,zhang2017lump}). Apart from taking long wave limits, one can also derive lump solutions via Hirota's method \cite{hirota2004direct,Hietarinta1997introduction} or the singular manifold method \cite{estevez2007algorithmic,albares2017lumps,weiss1983painleve}. Research has shown that lump solutions have many applications in nonlinear dynamics \cite{manukure2018lump}.  They provide appropriate prototypes to model rogue wave dynamics in oceanography \cite{muller2005rogue} and nonlinear optics \cite{solli2007optical}.

Recently, lump solutions have been found to generate a type of rogue wave solutions known as \textit{line rogue waves}, which may arise under some parameter constraint \cite{shi2017line}.  Line rogue waves \cite{shi2017line} usually emerge from a constant background with line profiles which eventually decay  into the constant background. Rogue waves have also been of considerable interest in recent years due to emerging applications in other contexts such as nonlinear optics \cite{solli2007optical} and the atmosphere \cite{stenflo2010rogue}. For rogue waves in other physical contexts, see \cite{onorato2013rogue}. The most universal mathematical model for the study of rogue waves is the one-dimensional focusing Nonlinear Schr\"odinger equation \cite{bertola2016rogue, clarkson2017rational}. However, other integrable models, notably, the Kadomtsev-Petviashvili equation \cite{ma2015lump,xu2014rogue}, the Hirota-Satsuma-Ito equation \cite{zhou2019lump,zhang2020dynamics} and the B-type KP equation \cite{gilson1990lump,feng2017rogue,yang2016lump} have also been used to study rogue waves. 

In this article, we study lump and line rogue wave solutions of a novel (2+1)-dimensional equation which is an extension of the so-called Hietarinta equation \cite{Hietarinta1997introduction}. We will employ Hirota's method \cite{hirota2004direct} which is perhaps the most effective tool for finding exact solutions, particularly soliton solutions (see e.g., \cite{ma2021n,ma2021soliton}), to nonlinear equations that possess Hirota bilinear forms. 
To use this method to find lump solutions, one constructs positive quadratic function solutions to a bilinear equation and uses logarithmic transformations to obtain the desired solutions (see e.g., \cite{ma2015lump,manukure20192+,ma2018lump,Batwa2018study,ma2020symbolic}). First, we introduce a (2+1)-dimensional equation as a modification of the (1+1)-dimensional Hietarinta equation. We further formulate its Hirota bilinear form and construct positive quadratic solutions to this bilinear equation. Consequently, we construct lump and line rogue waves to the newly introduced equation. The paper concludes with illustrative examples and some concluding remarks.

\section{A modified Hietarinta equation }
The bilinear Hietarinta equation \cite{Hietarinta1997introduction} is given by,
\begin{equation}
	(D_x^4-D_x D_t^3 +\alpha D_x^2 +\beta D_x D_t + \gamma D_t^2) f \cdot f=0,
\end{equation}
where $\alpha,\beta,\gamma$ are constants and $D_x, D_t$ are Hirota derivatives \cite{hirota2004direct}. This equation is in (1+1)-dimensions and is integrable in the sense that it has at least four soliton solutions and also passes the Painlev\'e test \cite{Hietarinta1997introduction,steeb1988nonlinear}. It is important to point out that there has been no report so far on the existence of rationally localized wave solutions to the Hietarinta equation. Recently, some (2+1)-dimensional extensions of the above equation have been introduced and shown to possess rationally localized wave solutions. For example, Batwa and Ma \cite{batwa2020lump} introduced the bilinear modified Hietarinta equation 
\begin{equation}
(\alpha_1D_x^4+\alpha_2D_x D_t^3+\gamma_1D_yD_t+\gamma_2D_x^2+\gamma_3 D_x D_t + \gamma_4D_xD_y+\gamma_5D_y^2) f \cdot f =0,
\end{equation}
and found lump solutions to the associated (2+1)-dimensional nonlinear equation. Manukure and Zhou \cite{manukure2021study} also introduced the modified Hietarinta bilinear equation  
\begin{equation}
(D_x^4+D_x D_t^3+\alpha D_x^2+\beta D_x D_t + \gamma D_t^2 -D_t D_y) f \cdot f =0,
\end{equation}
and derived lump and line rogue waves to the associated nonlinear equation. Here, we introduce yet another extension of the bilinear Hietarinta equation as follows,

\begin{equation}
\label{bilineareqn}
B(f)=(D_x^4+D_x D_t^3+\alpha D_x^2+\beta D_x D_t + \gamma D_t^2 -D_x D_y) f \cdot f =0. 
\end{equation}
Under the transformation,
\begin{equation}
\label{transformation}
u=2 ( \text{ln} f)_x, \qquad  v= 2 \text{ln} f,
\end{equation}
the corresponding (2+1)-nonlinear equation is found to be
\begin{equation}
\label{hietarintaeqn}
P(u)=6u_x u_{xx}+u_{xxxx}+3u_t u_{tt}+3u_{tx} v_{tt}+u_{xttt}+\alpha u_{xx}+\beta u_{tx}+\gamma u_{tt}-u_{xy}=0,
\end{equation}
where $v_x=u$, and $\alpha, \beta,$ and $\gamma$ are arbitrary constants. The direct connection between \eqref{bilineareqn} and \eqref{hietarintaeqn} is given by the equation
\begin{equation}
P(u)= \left( \frac{B(u)}{f} \right)_x.
\end{equation} 

To construct locally rationalized wave solutions, we use the method introduced in \cite{ma2018lump}. Thus, we find positive quadratic function solutions in the form
\begin{equation}
\label{formforf}
\begin{cases}
f&=g^2+h^2+a_9,\\
g&=a_1 x+a_2 y+ a_3 t + a_4, \\
h&=a_5 x +a_6 y +a_7 t +a_8.\\
\end{cases}
\end{equation}
where $a_i, 1\leq i\leq9$ are real and $a_9>0$. If we assume that $g$ and $h$ are linearly dependent, then $f$ can be reduced to
\begin{equation}
f=\delta_1(g+\delta_2)^2+\delta_3
\end{equation}
where $\delta_1, \delta_2, \delta_3$ are real and $\delta_1, ~\delta_3 > 0$.  It then follows that
\begin{equation} 
u=2(\ln f)_{x}=\frac{4a_1\delta_1(g+\delta_2)}{f}.
\end{equation}
The above solution is degenerate: 
\begin{equation}
\lim_{t\to \infty}u\ne 0,
\end{equation}
and for $t$ fixed, 
\begin{equation}
\lim_{x^2+y^2\to \infty}u\ne 0.
\end{equation}
When $g$ and $h$ are linearly independent, we will obtain lump solutions and line rogue waves under a certain parameter constraint. 

\section{Lump solutions}
Now, let us assume that $g$ and $h$ are linearly independent and the matrix

\begin{equation}\label{rank3} 
\left(\begin{array}{ccc}a_1 & a_2  \\a_5 & a_6 \end{array}\right)
\end{equation}
is of full rank. This means that the determinant of this matrix is nonzero, i.e.,
\begin{equation}
\label{deter}
\Delta := a_1 a_6 -a_2 a_5  \neq 0.
\end{equation}
Substituting $f$ in \eqref{formforf} into \eqref{bilineareqn}, we obtain the following solution set:
\begin{equation}
\begin{cases}
& a_2=\dfrac{\alpha a_1^3  + \beta a_1^2 a_3 + (\alpha a_5^2 +\gamma  a_3^2 -\gamma  a_7^2) a_1 +a_3 a_5 (\beta a_5 +2 \gamma a_7)}{a_1^2+a_5^2},\\
& a_6=\dfrac{\alpha a_5^3 + \beta a_5^2 a_7 + (\alpha a_1^2 -\gamma a_3^2  +\gamma a_7^2) a_5 +a_1 a_7 (\beta a_1 +2 \gamma a_3)}{a_1^2+a_5^2},\\
& a_9=-\dfrac{3(a_1^4+2a_1^2 a_5^2+a_3^3 a_1+a_3 a_7^2a_1+a_5 a_3^2 a_7 +a_5^4+a_5 a_7^3)(a_1^2+a_5^2)}{\gamma (a_1 a_7 - a_3 a_5)^2},\\
\end{cases}
\end{equation}
with the $a_i$'s as free parameters. To ensure the analyticity of the functions in \eqref{transformation}, we impose the conditions 
\begin{equation}\label{eqn0}
a_9>0,~~\gamma\neq0,~~	a_1 a_7 -a_3 a_5 \neq 0.
\end{equation}
Note that this condition is sufficient for condition \eqref{deter} since
\begin{equation}\label{const}
a_1 a_6 -a_2 a_5=\frac{(a_1 a_7 -a_3 a_5)(\beta a_1^2+\beta a_5^2+2\gamma a_1a_3+2\gamma a_5a_7)}{a_1^2+a_5^2}.
\end{equation}
Thus, positive quadratic function solutions to the bilinear modified Hietarinta equation \eqref{bilineareqn} take the form:
\begin{equation}
\begin{split}
\label{solf}
f=&(a_1 x + \frac{\alpha a_1^3  + \beta a_1^2 a_3 + (\alpha a_5^2 +\gamma  a_3^2 -\gamma  a_7^2) a_1 +a_3 a_5 (\beta a_5 +2 \gamma a_7)}{a_1^2+a_5^2} y + a_3 t +a_4)^2 \\
&+(a_5 x +\frac{\alpha a_5^3 + \beta a_5^2 a_7 + (\alpha a_1^2 -\gamma a_3^2  +\gamma a_7^2) a_5 +a_1 a_7 (\beta a_1 +2 \gamma a_3)}{a_1^2+a_5^2} y + a_7 t +a_8)^2 \\
&-\frac{3(a_1^4+2a_1^2 a_5^2+a_3^3 a_1+a_3 a_7^2a_1+a_5 a_3^2 a_7 +a_5^4+a_5 a_7^3)(a_1^2+a_5^2)}{\gamma (a_1 a_7 - a_3 a_5)^2},
\end{split}
\end{equation}
which consequently yield the following solution to the modified Hietarinta equation \eqref{hietarintaeqn},
\begin{equation}\label{solu}
u=\frac{4a_5h+4a_1 g}{f},
\end{equation}
under the transformation \eqref{transformation}, with $g$ and $h$ given by,
\begin{equation}
\begin{cases}
g&=a_1 x + \dfrac{\alpha a_1^3  + \beta a_1^2 a_3 + (\alpha a_5^2 +\gamma  a_3^2 -\gamma  a_7^2) a_1 +a_3 a_5 (\beta a_5 +2 \gamma a_7)}{a_1^2+a_5^2} y + a_3 t +a_4, \\
h&=a_5 x +\dfrac{\alpha a_5^3 + \beta a_5^2 a_7 + (\alpha a_1^2 -\gamma a_3^2  +\gamma a_7^2) a_5 +a_1 a_7 (\beta a_1 +2 \gamma a_3)}{a_1^2+a_5^2} y + a_7 t +a_8. 
\end{cases}
\end{equation}
The solution \eqref{solu} satisfies the condition
\begin{equation}\label{lumpdec}
\lim\limits_{x^2+y^2\to\infty}u(x,y,t)=0
\end{equation}
for any fixed $t$ and is therefore a lump solution to equation \eqref{hietarintaeqn}. 



\section{Rogue waves}
Rogue waves are large oceanic waves which are localized in both time and space \cite{ward2019evaluating,liu2018rogue}. In this section, we find rogue wave solutions to the modified Hietarinta equation \eqref{hietarintaeqn} by requiring the matrix \eqref{rank3} to be rank deficient. This requirement imposes a constraint on some of the parameters in \eqref{const} which consequently yields line rogue waves. Line rogue waves are known to emerge from a constant background and eventually disappear into the same background. 

Now, suppose again that $g$ and $h$ are linearly independent and 
\begin{equation}\label{rank4} 
\mathrm{rank}\left(\begin{array}{ccc}a_1 & a_2  \\a_5 & a_6 \end{array}\right)<2. 
\end{equation}
Let $\eta(x,y)=a_1x+a_2y$. Then, we have
\[ g=\eta+a_3 t+a_4,\quad h=\delta\eta+a_7t+a_8, \]  
for some $\delta\in \mathbb{R}$. It follows that $f$ can be written in the form
\[ f=(\delta_1\eta +\delta_2t+\delta_3)^2+(\delta_4t+\delta_5)^2+\delta_6, \]
where $\delta_1,\delta_2,\cdots, \delta_6$ are real constants and $\delta_1\ne 0,\delta_4\ne 0,\delta_6>0.$ 

From the rank condition \eqref{rank4}, we have
\begin{equation*}
a_1 a_6 -a_2 a_5=\frac{(a_1 a_7 -a_3 a_5)(\beta a_1^2+\beta a_5^2+2\gamma a_1a_3+2\gamma a_5a_7)}{a_1^2+a_5^2}=0,
\end{equation*}
which gives rise to the constraint,
\begin{equation}\label{cond}
\beta a_1^2+\beta a_5^2+2\gamma a_1a_3+2\gamma a_5a_7=0
\end{equation}
as a result of condition \eqref{eqn0}. According to \eqref{eqn0}, at least one of the constants, $a_1,~a_5$ is nonzero. If we assume that $a_5\neq0$, we can rewrite the above condition \eqref{cond} as,
\begin{equation}\label{cond1}
a_7=-\frac{\beta a_1^2+\beta a_5^2+2\gamma a_1a_3}{2\gamma a_5},
\end{equation}
for $\gamma\neq0$. Under this condition \eqref{cond1} and condition \eqref{eqn0}, the solutions in \eqref{solu} yield a class of solutions that satisfy
\begin{equation}\label{cond2}
\lim\limits_{|t|\to\infty}u(x,y,t)=0
\end{equation}
for $(x,y)\in\mathbb{R}^2$ uniformly. This shows that the resulting solutions are localized in time.

\section{Illustrative examples}
To depict the dynamical behavior of the localized wave solutions, we choose certain specific values for the parameters. 
\subsection{Lump solutions}

Choosing the parameters, 
$$ \alpha=1, \beta=-1, \gamma=1, a_1=-1, a_3=2,a_4=-2,a_5=1,a_7=-3,a_8=3,$$
we obtain the nonlinear equation
\begin{equation}\label{nonl1}
6u_x u_{xx}+u_{xxxx}+3u_t u_{tt}+3u_{tx} v_{tt}+u_{xttt}+ u_{xx}-u_{tx}+ u_{tt}-u_{xy}=0,
\end{equation}
with corresponding bilinear equation 
\begin{equation}
(D_x^4+D_x D_t^3+ D_x^2- D_x D_t + D_t^2 -D_x D_y) f \cdot f =0. 
\end{equation}
The quadratic function solutions to the above bilinear equation is given by
\begin{equation}
	f= \left(2t-x-\frac{13}{2}y-2\right)^2+\left(-3t+x+\frac{25}{2}y+3\right)^2+366,
\end{equation}
and the corresponding lump solutions to the nonlinear equation \eqref{nonl1} is
\begin{equation}\label{lumpsol}
	u=-\frac{8(5t-2x-19y-5)}{26t^2-20tx-202ty+4x^2+76xy+397y^2-52t+20x+202y+758}.
\end{equation}
If we choose the the values $t=-20,~0$ and $20$, we get the particular solutions 
\begin{equation}\label{lump4}
	u=\frac{8(105+2x+19y)}{4x^2+76xy+397y^2+420x+4242y+12198},
\end{equation}
\begin{equation}\label{lump5}
	u=\frac{8(5+2x+19y)}{4x^2+76xy+397y^2+20x+202y+758},
\end{equation}
and
\begin{equation}\label{lump6}
	u=\frac{8(-95+2x+19y)}{4x^2+76xy+397y^2-380x-3838y+10118},
\end{equation}
respectively, with 3D and contour plots shown below. 

One can easily verify that the above solutions $u$ decays in all spacial directions, ie., they satisfy the condition \eqref{lumpdec}.

	\begin{figure}
		\centering
			\subfloat{\includegraphics[width=6.0cm]{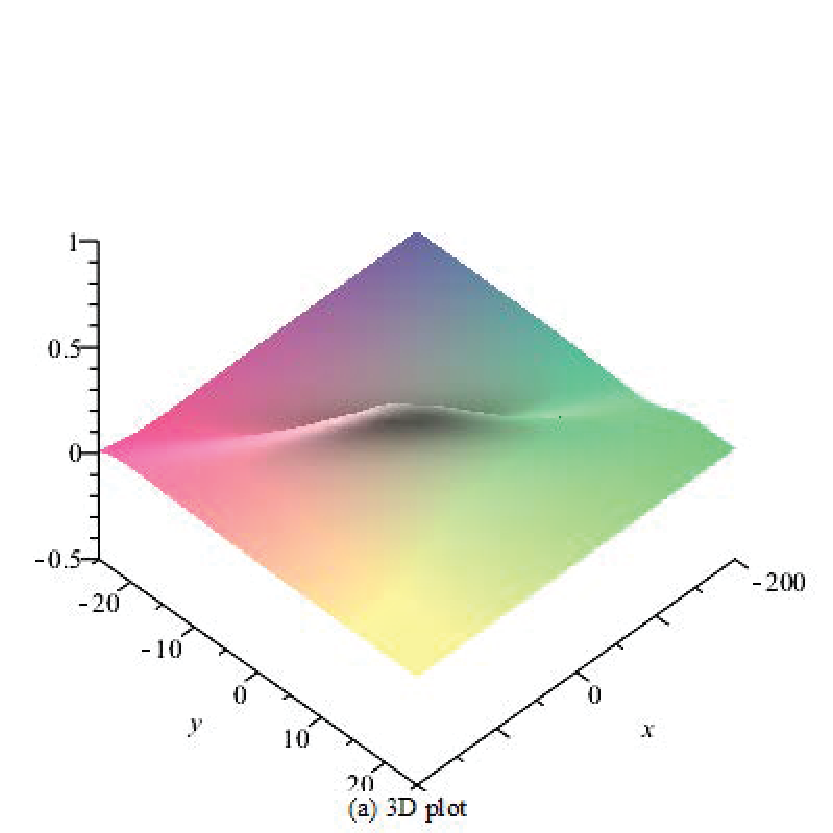}}
			\subfloat{\includegraphics[width=6.0cm]{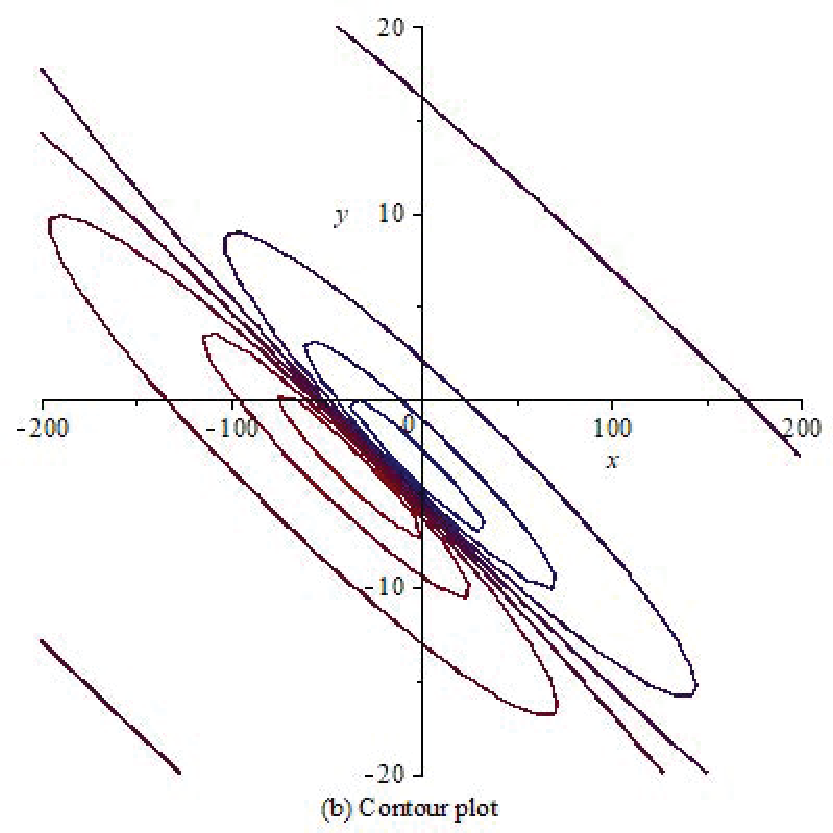}}
		\caption{Wave profile of solution \ref{lump4}\\}\label{fig:4}
	\end{figure}
	\begin{figure}
		\centering
		\subfloat{\includegraphics[width=6.0cm]{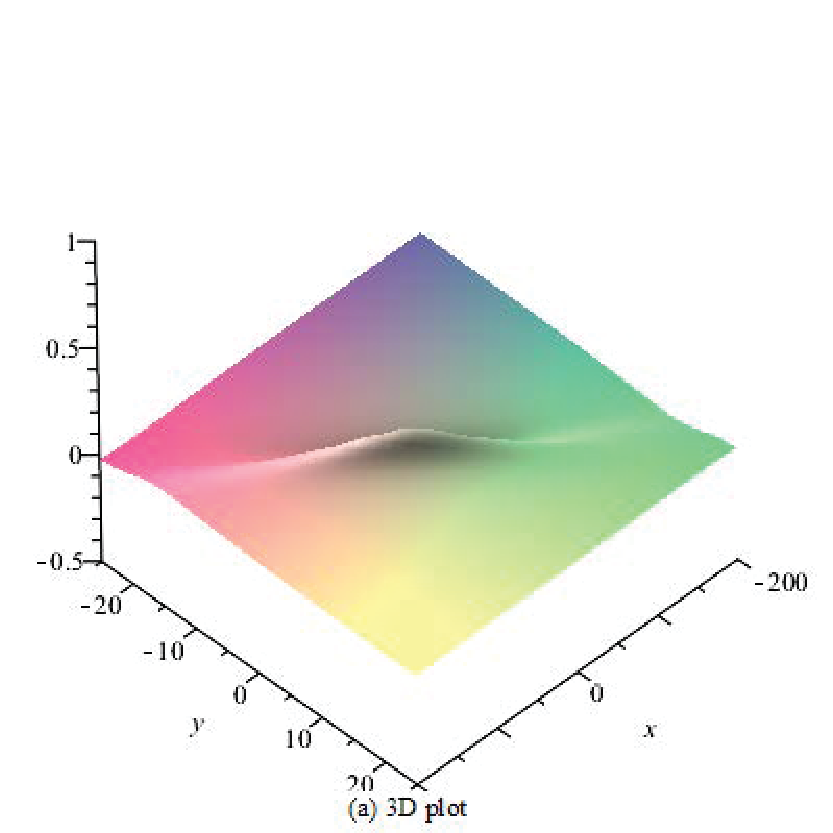}}
	\subfloat{\includegraphics[width=6.0cm]{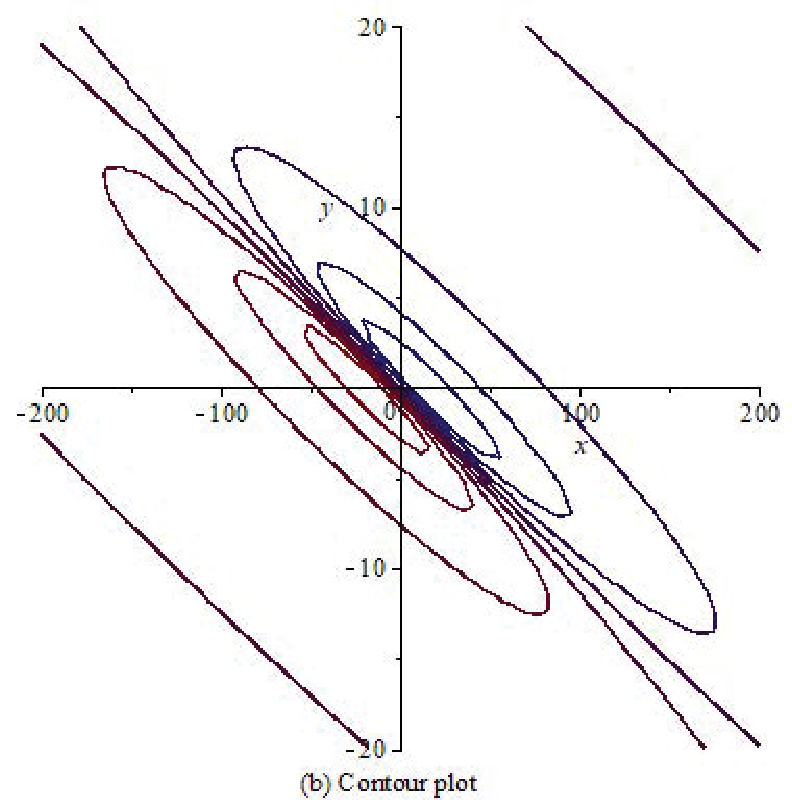}}
		\caption{Wave profile of solution \ref{lump5}\\}\label{fig:5}
	\end{figure}
	\begin{figure}
		\centering
		\subfloat{\includegraphics[width=6.0cm]{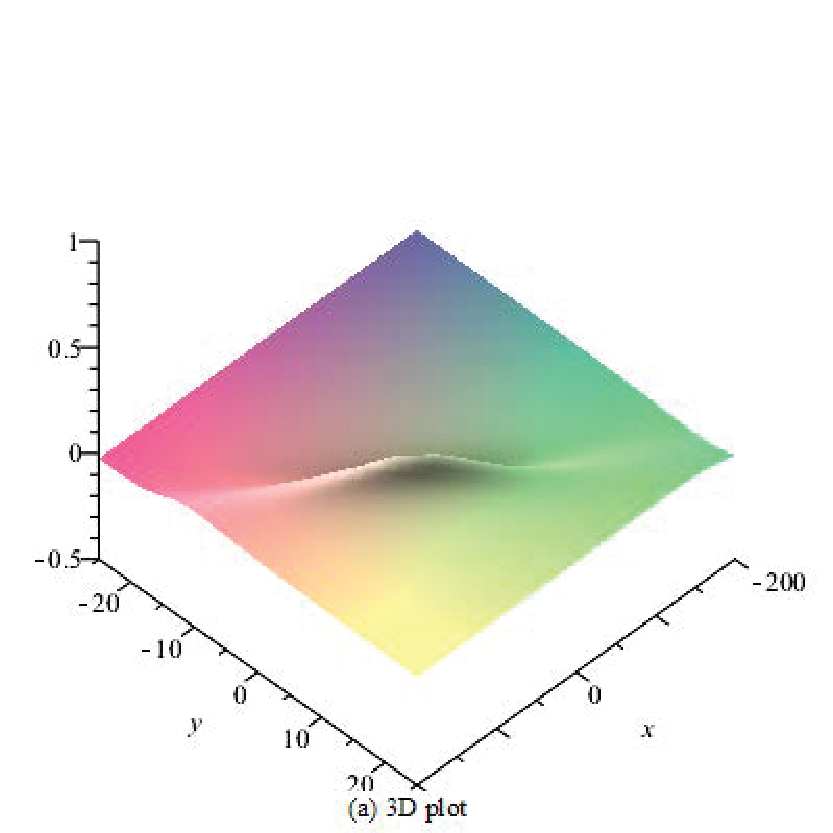}}
		\subfloat{\includegraphics[width=6.0cm]{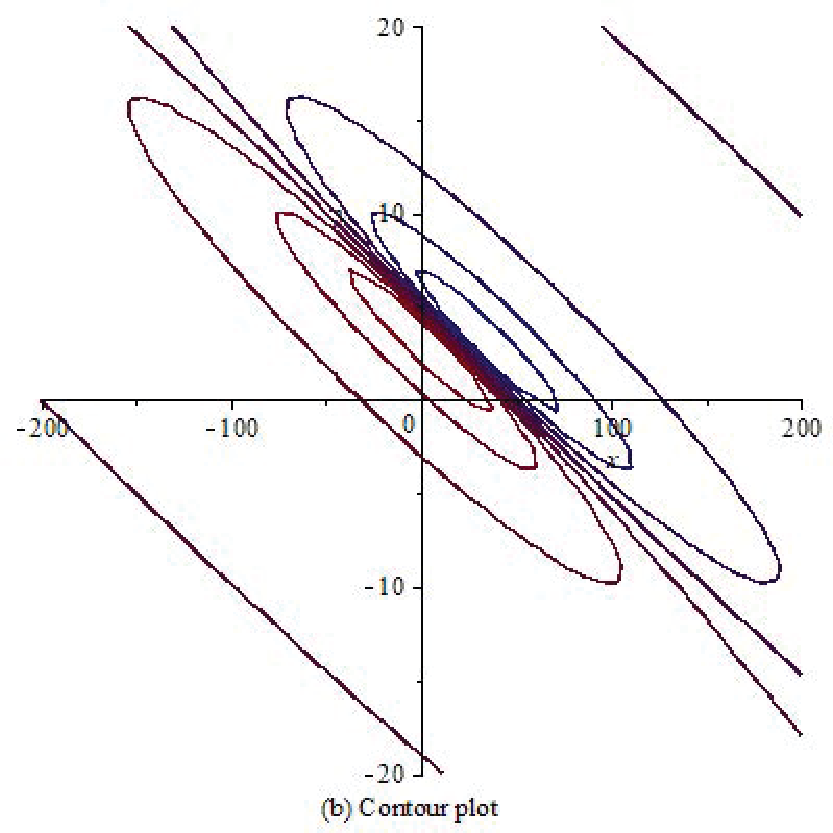}}
		\caption{Wave profile of solution \ref{lump6}\\}\label{fig:6}
	\end{figure}

The amplitude of the wave function \eqref{lumpsol} is $\displaystyle\frac{2}{\sqrt{283}}$ which is the height of the wave for all values of $t$. Thus, the lump solutions propagate with a constant amplitude at all times.

 \subsection{Line rogue waves}
For line rogue waves, we must choose parameters that satisfy not only condition \eqref{eqn0}, but also condition \eqref{cond}. 
To this end, we choose
$$ \alpha=1, \beta=1, \gamma=-1, a_1=1, a_3=2,a_4=0,a_5=-1,a_7=1,a_8=3=0.$$
Consequently, we obtain the nonlinear equation
\begin{equation}\label{nonl2}
6u_x u_{xx}+u_{xxxx}+3u_t u_{tt}+3u_{tx} v_{tt}+u_{xttt}+ u_{xx}+ u_{tx}- u_{tt}-u_{xy}=0, 
\end{equation}
with corresponding bilinear equation
\begin{equation}
(D_x^4+D_x D_t^3+D_x^2+ D_x D_t + - D_t^2 -D_x D_y) f \cdot f =0. 
\end{equation}
The positive quadratic function solution to the above bilinear equation is given by
\begin{equation}
	f= \left(2t+x+\frac{7}{2}y\right)^2+\left(t-x-\frac{7}{2}y\right)^2+6,
\end{equation}
and the corresponding lump solution to the nonlinear equation \eqref{nonl2} is  
\begin{equation}\label{rogsol}
	u=\frac{8(t+2x+7y)}{10t^2+4tx+14ty+4x^2+28xy+49y^2+12}.
\end{equation}
If we let $t=-10,-4,-2,0, 2, 4$ and $10$, we obtain the particular solutions, 

\begin{equation}\label{rog0}
	u=\frac{8(-10+2x+7y)}{4x^2+28xy+49y^2-40x-140y+1012},
\end{equation}
\begin{equation}\label{rog1}
	u=\frac{8(-4+2x+7y)}{4x^2+28xy+49y^2-16x-56y+172},
\end{equation}
\begin{equation}\label{rog2}
	u=\frac{8(-2+2x+7y)}{4x^2+28xy+49y^2-8x-28y+52},
\end{equation}
\begin{equation}\label{rog3}
	u=\frac{8(2x+7y)}{4x^2+28xy+49y^2+12},
\end{equation}
\begin{equation}\label{rog4}
	u=\frac{8(2+2x+7y)}{4x^2+28xy+49y^2+8x+28y+52},
\end{equation} 
\begin{equation}\label{rog5}
	u=\frac{8(4+2x+7y)}{4x^2+28xy+49y^2+16x+56y+172},
\end{equation}
and 
\begin{equation}\label{rog6}
	u=\frac{8(10+2x+7y)}{4x^2+28xy+49y^2+40x+140y+1012},
\end{equation}
respectively. The 3D plot and contour plots for these solutions are shown below.\\ 
\begin{figure}
	\centering
		\subfloat{\includegraphics[width=6.0cm]{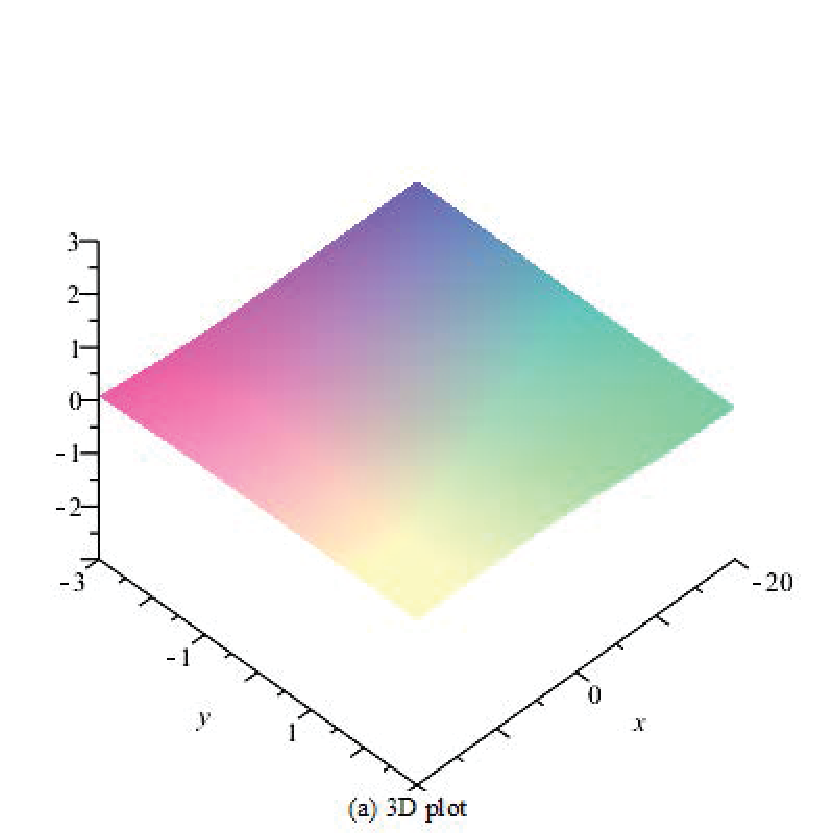}}
		\subfloat{ \includegraphics[width=6.0cm]{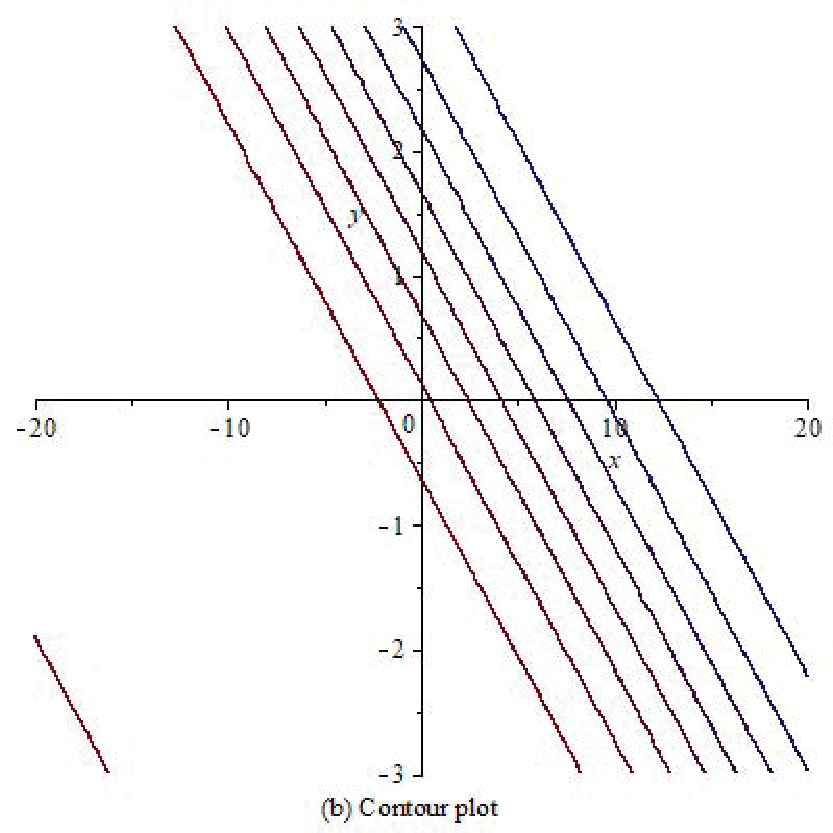}}
	\caption{\\Wave profile of solution \ref{rog0}}\label{fig:7}
\end{figure}

\begin{figure}
	\centering
		\subfloat{\includegraphics[width=6.0cm]{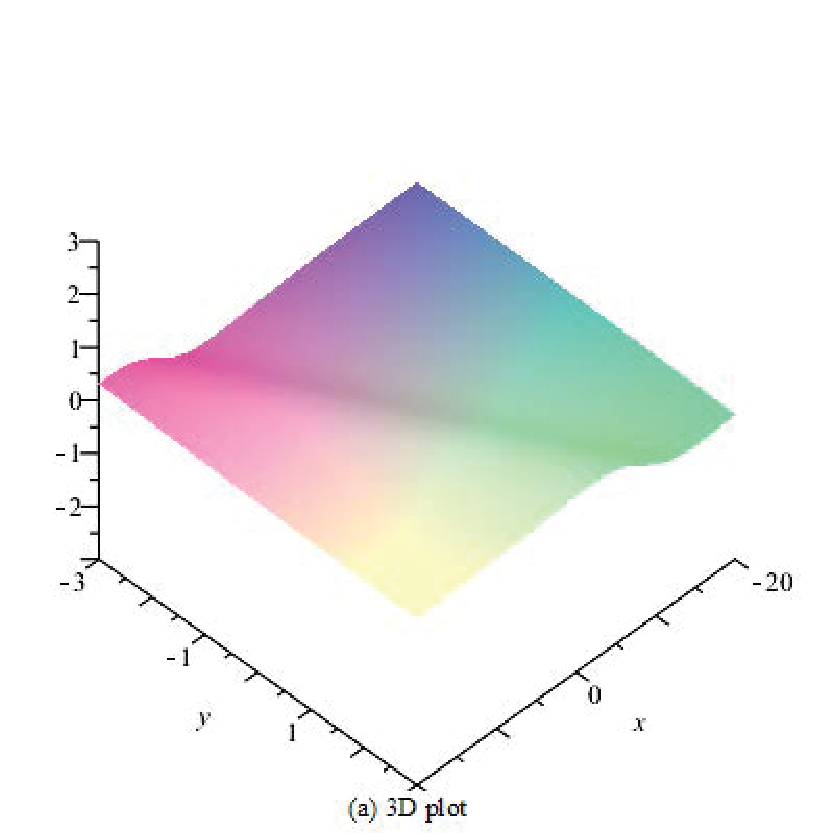}}
		 \subfloat{\includegraphics[width=6.0cm]{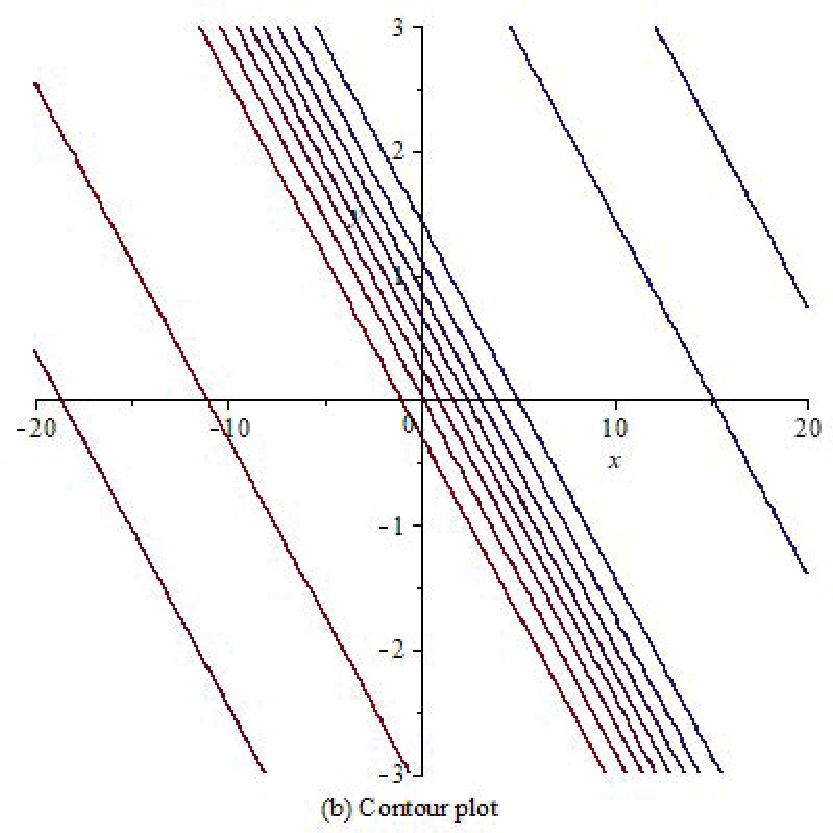}}
	\caption{\\Wave profile of solution \ref{rog1}}\label{fig:8}
\end{figure}

\begin{figure}
	\centering
	\subfloat{\includegraphics[width=6.0cm]{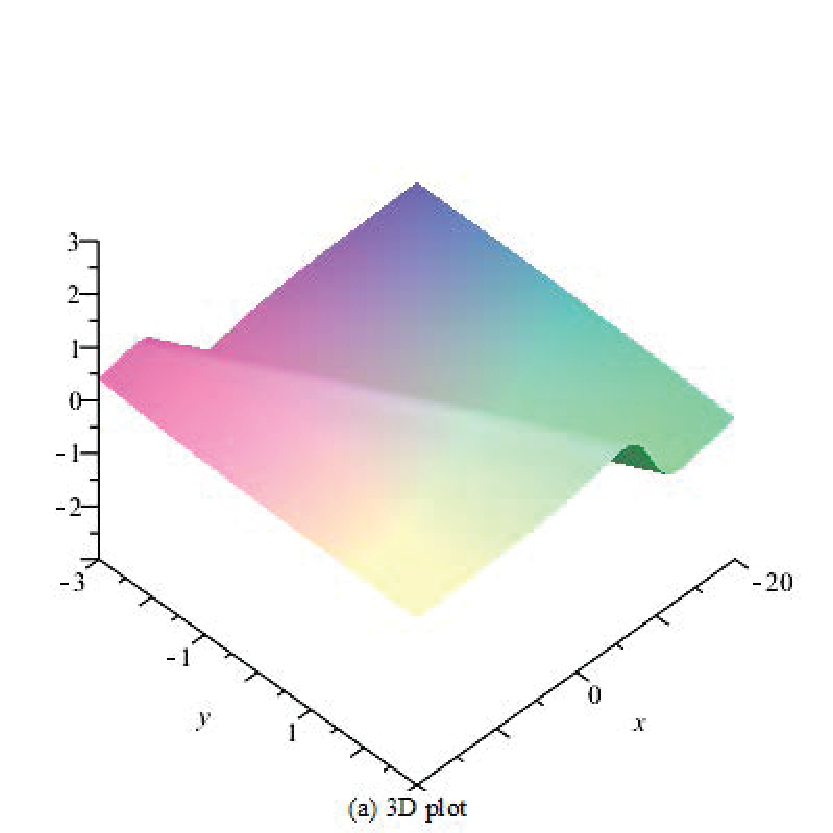}}
	\subfloat{\includegraphics[width=6.0cm]{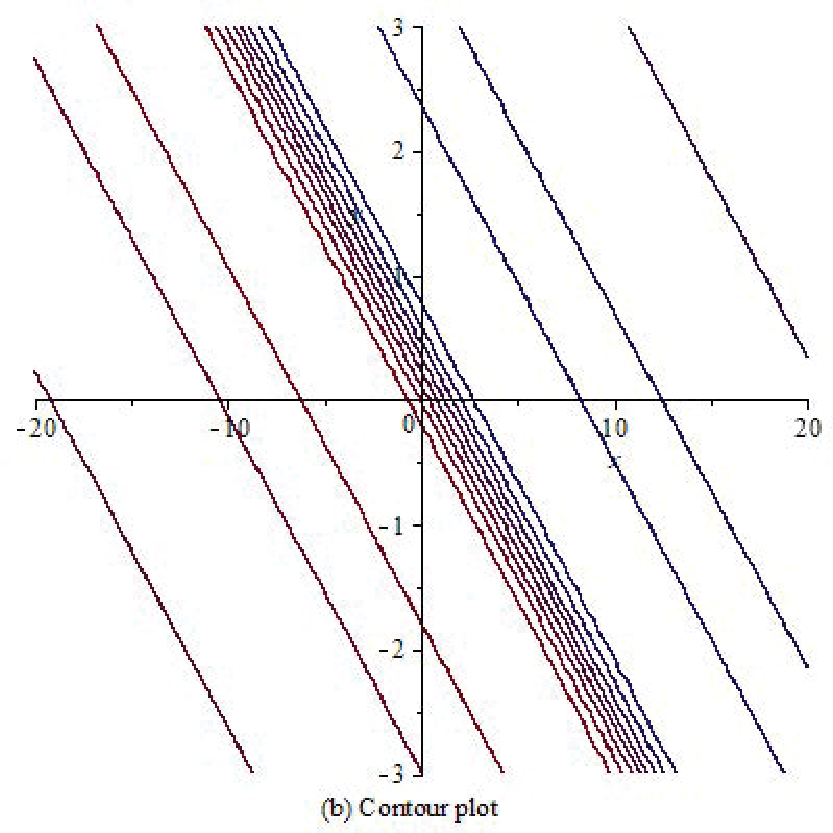}}
	\caption{\\Wave profile of solution \ref{rog2}}\label{fig:9}
\end{figure}

\begin{figure}
	\centering
	\subfloat{\includegraphics[width=6.0cm]{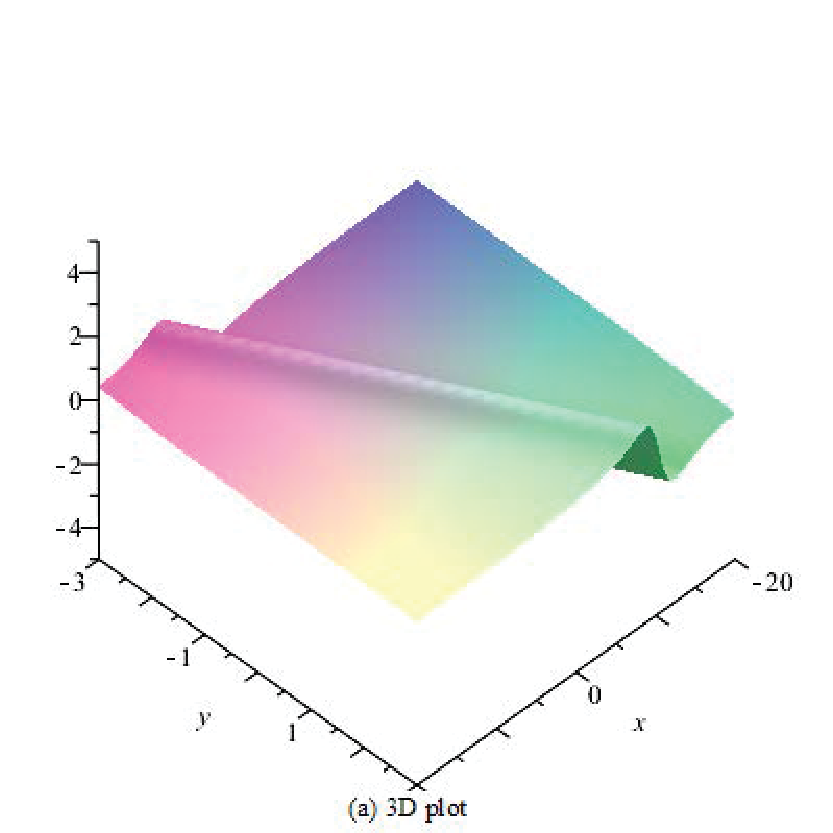}}
	\subfloat{\includegraphics[width=6.0cm]{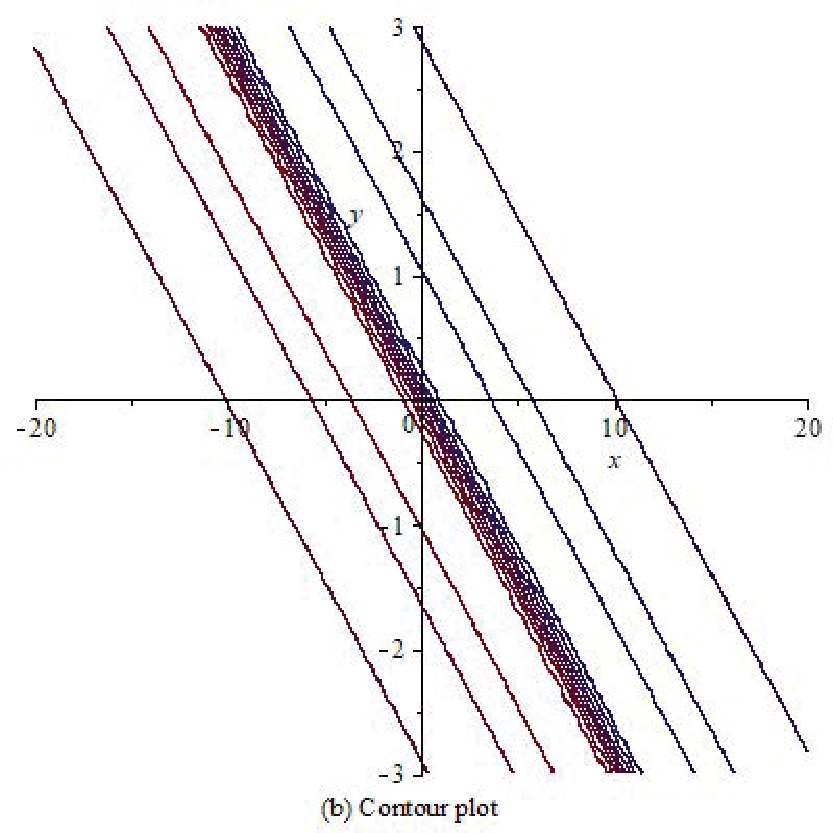}}
	\caption{\\Wave profile of solution \ref{rog3}}\label{fig:10}
\end{figure}

 \begin{figure}
 	\centering
	\subfloat{\includegraphics[width=6.0cm]{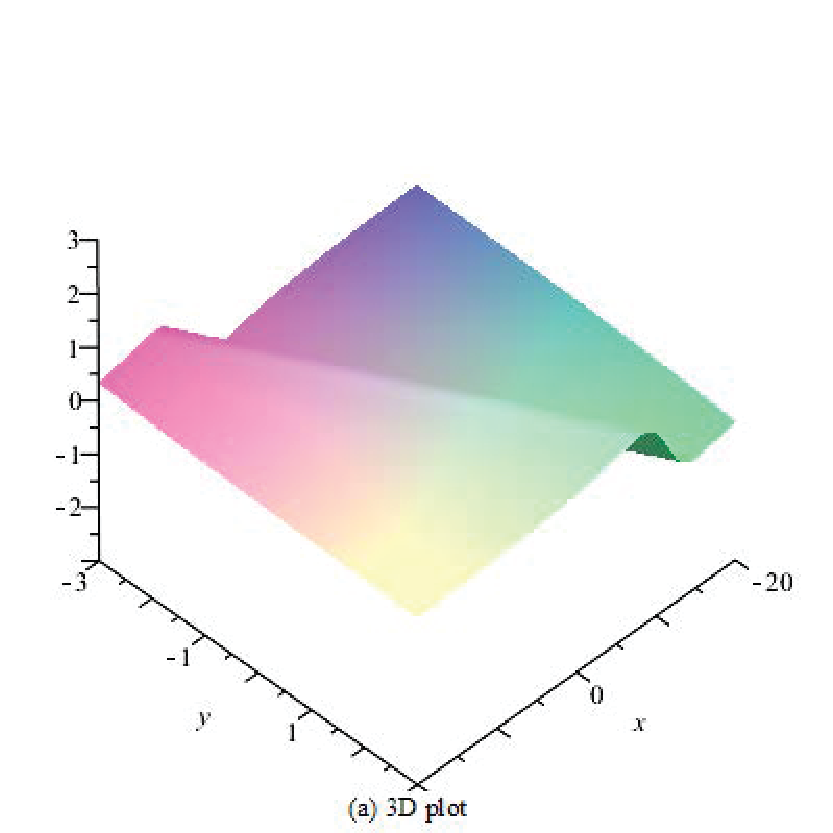}}
	\subfloat{\centering \includegraphics[width=6.0cm]{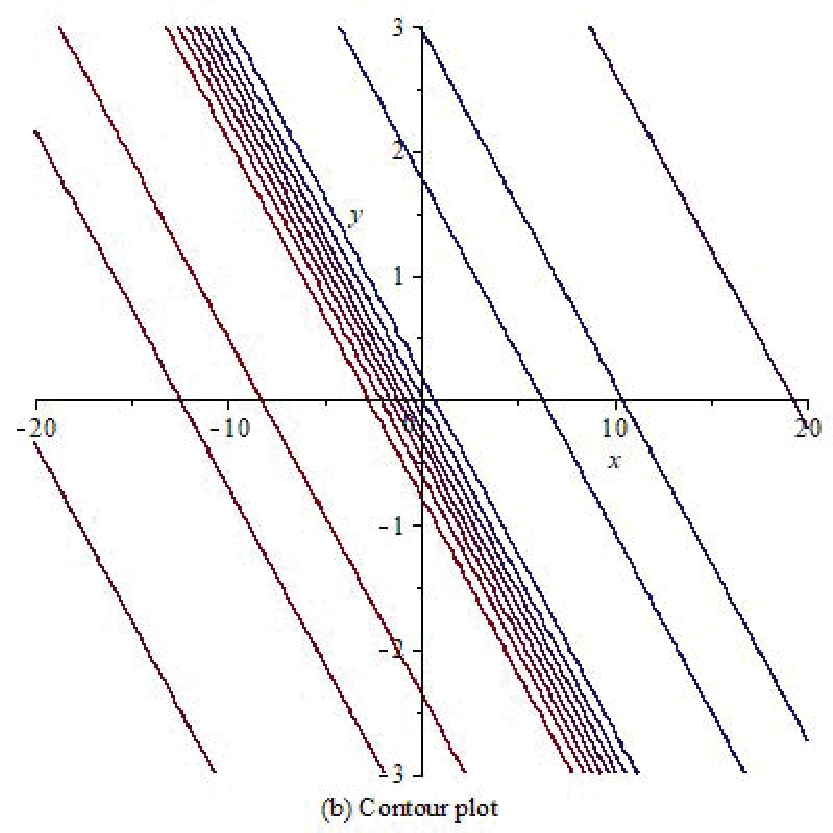}}
	\caption{\\Wave profile of solution \ref{rog4}\\}\label{fig:11}
\end{figure}

\begin{figure}
	\centering
		\subfloat{ \includegraphics[width=6.0cm]{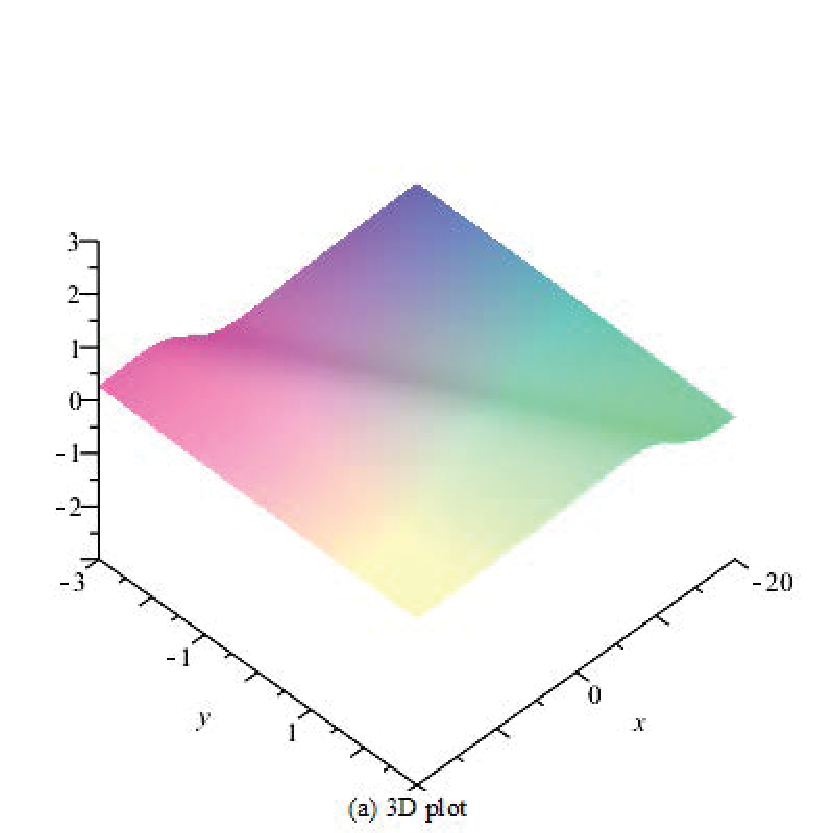}}
		\subfloat{\includegraphics[width=6.0cm]{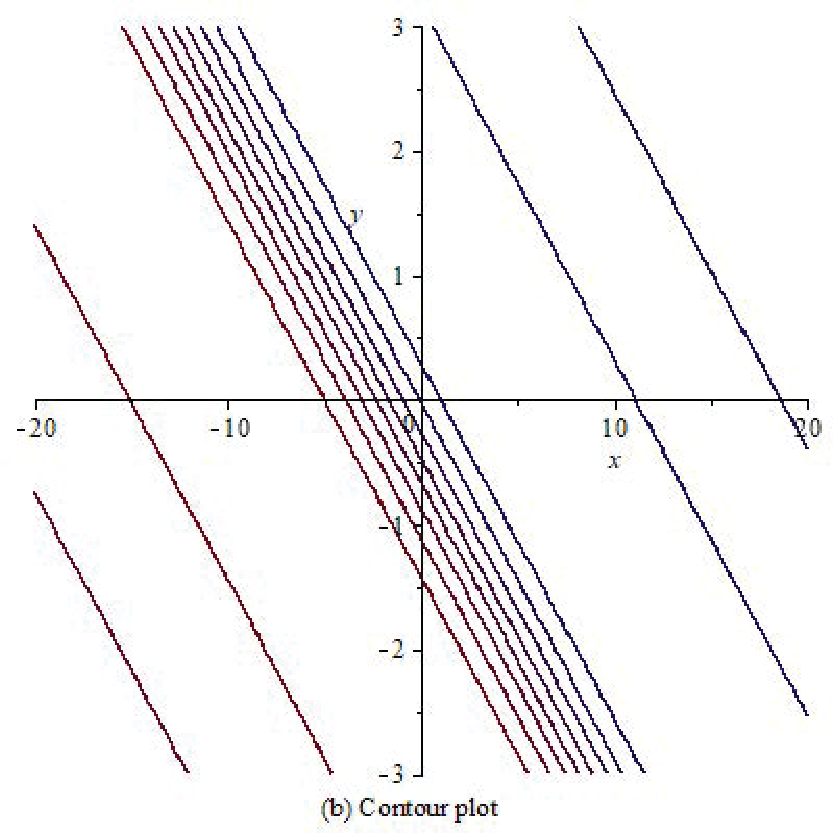}}
		\caption{Wave profile of solution \ref{rog5}\\}\label{fig:12}
	\end{figure}

\begin{figure}
	\centering
		\subfloat{\includegraphics[width=6.0cm]{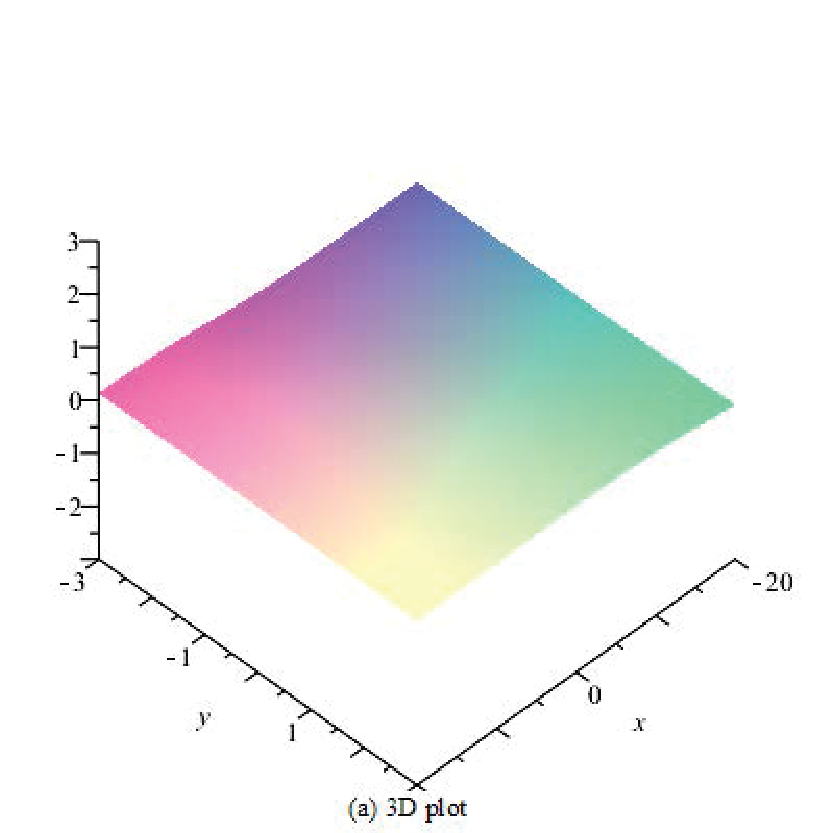}}
		\subfloat{ \includegraphics[width=6.0cm]{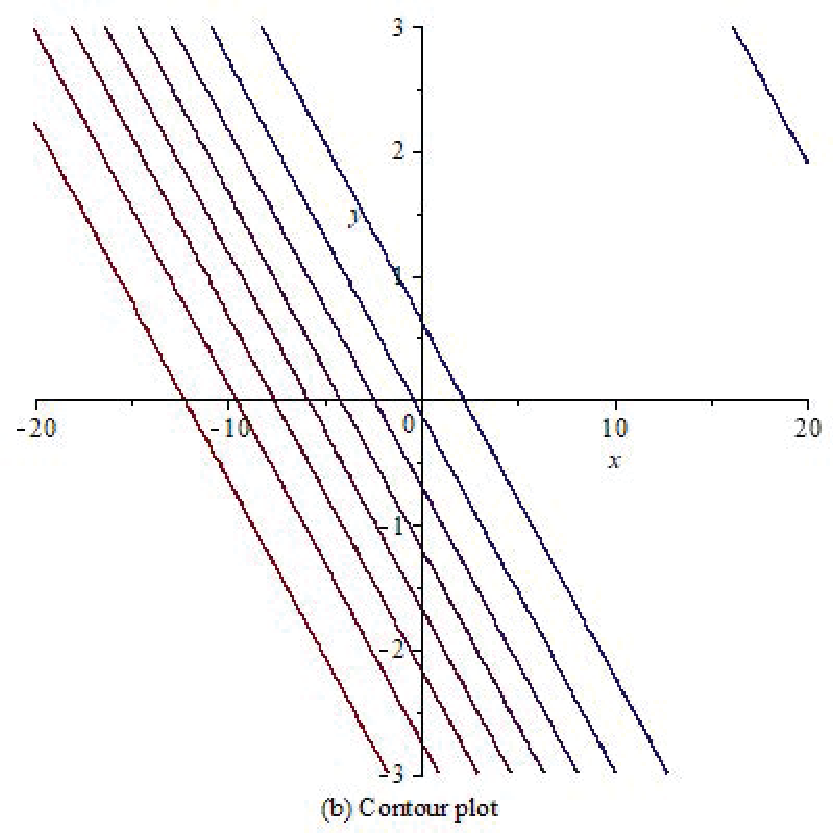}}
	\caption{Wave profile of solution \ref{rog6}\\}\label{fig:13}
\end{figure}

The solution $u$ in \eqref{rogsol} satisfies condition \eqref{cond2} as well as
\begin{equation}
\lim_{x^2+y^2+t^2\to\infty} u(t,x,y)=0,
\end{equation}
unless $2x+7y=\epsilon$, for any $\epsilon\in \mathbb{R}$. The maximum of this solution \eqref{rogsol} is $\displaystyle\frac{2}{\sqrt3}$ which occurs when $t=0$. This means the solution rises from the constant background, peaks when $t=0$ (see Fig. 7) and eventually decays into the constant background. 

  \section{Concluding Remarks}
By means of the Hirota bilinear method, we have constructed lump and rogue wave solutions to a so-called modified Hietarinta equation formulated from the (1+1)-dimensional Hietarinta equation. The lump solutions arise from quadratic function solutions of the associated bilinear equation whereas the rogue waves arise from a certain parameter constraint. The lump solutions have been shown to be spatially localized while the line rogue waves are both spatially and temporally localized. More specifically, the lump solutions are localized in all spatial directions and propagate with a constant amplitude of $2/\sqrt{283}$ for all values of $t$ on a constant background.  The line rogue wave on the other hand emerges with a line profile from a constant background and rises in amplitude or height to a maximum of $2/\sqrt{3}$ after which it begins to decay and finally disappears into the constant background.

As indicated earlier, a few other modifications of the Hietarinta equation \cite{Hietarinta1997introduction} have been presented in literature \cite{batwa2020lump,manukure2021study}. The modified Hietarinta equation presented by Batwa and Ma in \cite{batwa2020lump} possesses only lump solutions. 
In the case of Manukure and Zhou \cite{manukure2021study}, two classes of lump solutions and two classes of line rogue waves were found. We suspect therefore that the existence of line rogue waves in the equation presented in the current paper and the one in \cite{manukure2021study} may be due to the presence of the term $u_{tt}$ or $D^2_t$ which is missing in the Batwa-Ma equation. Our equation therefore adds to the list of examples of nonlinear partial differential equations which possess lump solutions and line rogue waves.


\bibliography{Hietarinta}
\bibliographystyle{unsrt}

\end{document}